\documentclass[fleqn,12pt,twoside]{article}
\usepackage{espcrc1}

\usepackage{graphicx}
\usepackage[figuresright]{rotating}

\newcommand{\beq}{\begin{eqnarray}}
\newcommand{\eeq}{\end{eqnarray}}

\newcommand{\bfm}{\mbox{{\boldmath $m$}}}
\newcommand{\la}{\langle}
\newcommand{\ra}{\rangle}

\newcommand{\AmS}{{\protect\the\textfont2
  A\kern-.1667em\lower.5ex\hbox{M}\kern-.125emS}}

\title{Effects of Vector Coupling on Chiral and Color-superconducting
       Phase Transitions -- interplay among the scalar, pairing and
       vector interaction --}

\author{M. Kitazawa\address[KYOTO]{Department of Physics,
 Kyoto University, Kyoto 606-8502, Japan},
T. Koide\address[YITP]{Yukawa Institute for Theoretical Physics,
Kyoto University, Kyoto 606-8502, Japan},
T. Kunihiro\addressmark[YITP]
and
Y. Nemoto\address[BNL]{RIKEN BNL Research Center, BNL, Upton, NY 11973}}

\begin{document}

% typeset front matter
\maketitle

\begin{abstract}
We investigate effects of the 
vector interaction appearing in chiral effective Lagrangians
 on the chiral and color superconducting (CSC) 
phase transitions using Nambu-Jona-Lasinio model.
It is  shown that the repulsive density-density interaction 
coming from the vector term enhances 
 competition between the chiral symmerty breaking ($\chi$SB)
and CSC phase transition:
When the vector coupling is increased,
the first order transition
between the $\chi$SB and CSC phase becomes weaker, 
and the coexisting phase in which both the chiral and 
color-gauge symmetries are dynamically broken comes 
to exisit in a wider region in the $T$-$\mu$ plane.
We find that the critical line of
 the first order transition can have {\em two} endpoints
for an intermediate range of the vector coupling.
\end{abstract}

\section{Introduction}

It is one of the central issues in hadron physics
to determine the  phase diagram of 
strongly interacting matter in finite temperature $T$
and density.
Recent renewed interst in color superconductivity (CS) \cite{ref:CSC}
has stimulated intensive studies in this field,
which have been revealing a rich phase structure at finite density
quark matter \cite{ref:REV}.

%Now it is widely believed on the basis of effective theories 
%\cite{ref:REV}
%that the phase transition from the chiral symmetry breaking ($\chi$SB) 
%to the color superconducting (CSC) phase
%is first order at lower temperature,
%and this critical line continues for higher temperature 
%in the $T$-$\mu$ plane and ends at some point which is called 
%the endpoint.  

However, although  many works
on CS have been carried out with the use of effective chiral models, 
an important interaction in the vector channel
\beq
 {\cal L}_V=-G_V (\bar{\psi} \gamma ^{\mu}\psi)^2 
\eeq
has been scarcely considered.
%, which is found to alter the nature of the transition 
%from $\chi$SB to CS phase drastically, as we shall show.
The purpose of this article is to reveal new characteristics 
of the chiral to CSC transition based on 
a simple effective model incorporating ${\cal L}_V$
and explore how the vector interaction affect 
the phase structure\cite{ref:K3N}.

The significance to incorporate ${\cal L}_V$ may be understood as follows.
First of all, 
the instanton-anti-instanton molecule model, 
as well as the renormalization-group equation, 
show that ${\cal L}_V$ appears as a part of the effective 
Lagrangians.
Furthermore, 
since the vector interaction ${\cal L}_V$ includes the term 
$(\bar{\psi}\gamma^0 \psi)^2$, 
it gives rise to a repulsive energy proportional to the density squared.
Notice that the chiral transition at finite density is 
necessarily accompanied by a change in quark densty.
Then one expects naturally that ${\cal L}_V$
causes large effects on the chiral transition;
in fact, ${\cal L}_V$ is known to
 weaken the chiral phase 
transition and postpone the transition to a larger chemical
potential $\mu$.

\section{Thermodynamic Potential}

To investigate the effects of the vector interaction,
we use a simple Nambu-Jona-Lasinio model
with two flavors and three colors,
\begin{eqnarray}
{\cal L}&=& \bar{\psi}(i\gamma\cdot \partial -\bfm)\psi + {\cal L}_V\nonumber\\
&& + G_S \big\{ (\bar{\psi}\psi)^2
+ (\bar{\psi} i \gamma_5 {\bf \tau } \psi)^2 \big\} % \nonumber\\
+ G_{C} % \big\{ 
(\bar{\psi} i \gamma_5 \tau_2 \lambda_2 \psi^C )
(\bar{\psi}^C i \gamma_5 \tau_2 \lambda_2 \psi).
%+ (\bar{\psi} \tau_2 \lambda_2 \psi^C )
%(\bar{\psi}^C \tau_2 \lambda_2 \psi) \big\}.
\end{eqnarray}
Here, $\bfm =\mbox{diag}(m_u,m_d)$ 
is the current quark mass of up and down quarks,
$\psi^{C}\equiv C\bar{\psi}^{T}$, with 
$C=i\gamma^2\gamma^0$ being the charge conjugation operator, and 
$\tau_2$ and $\lambda_2$ are the second component 
of the Pauli  and the Gell-Mann matrices, representing 
the flavor SU$(2)_F$ and the color SU$(3)_C$, respectively.

The scalar coupling constant $G_S$ 
and the three momentum cutoff $\Lambda$ are chosen 
so as to reproduce the pion mass $m_{\pi}$ and the pion 
decay constant $f_{\pi}$ with 
the current quark mass $m_u=m_d=5.5$ MeV.
The $G_C/G_S=0.6$ is adopted, which gives a similar results
for CSC obtained with the instanton-induced interaction \cite{ref:BR}.
As for the vector coupling,
we shall vary it as a free parameter within a range given 
in the literature to clarify
 the effects of the vector coupling on the phase diagram.

To determine the phase diagram,
we shall calculate the thermodynamic potential $\Omega$
in  the mean-field approximation(MFA),
assuming the chiral  condensate, 
$ M_D=-2G_S\la\bar\psi\psi\ra $ and diquark condensate,
$ \Delta = \la \bar{\psi}^C i\gamma_5 \tau_2 \lambda_2 \psi \ra $.
The optimal values of these condensates  at given $T$ and $\mu$ are
determined so that 
$\Omega$ has the minimum value.
%we assume the rotational invariance of the system,
%which implies that only the 0-th component of the vector interaction 
%has an vacuum expectation value.
We notice that  ${\cal L}_V$ in MFA reduces to
$ -2G_V\bar\psi\gamma^0\psi \la\bar\psi\gamma^0\psi\ra 
+ G_V \la\bar\psi\gamma^0\psi\ra^2 $.
Owing to the first term,
it is found convenient to introduce  a shifted chemical potential  as
$ \tilde\mu = \mu - 2G_V\rho_q $.
% similar to the $\sigma$-$\omega$ model.
Here, $ \rho_q \equiv \la\bar\psi\gamma^0\psi\ra $
is the quark number density: In the following,
we shall show the results in terms of
the baryon density $\rho_B=1/3\cdot \rho_q$
 and the baryon chemical potential $\mu_B=3\mu$.
%which is determined by the thermodynamic relation
%$ \rho_q = -\partial\omega/\partial\mu $.

\section{Phase Diagrams}

As preliminary to the discussion 
on the effects of the vector interaction, 
we first present the phase structure without the vector interaction.
\begin{figure}
\begin{center}
\includegraphics[scale=1]{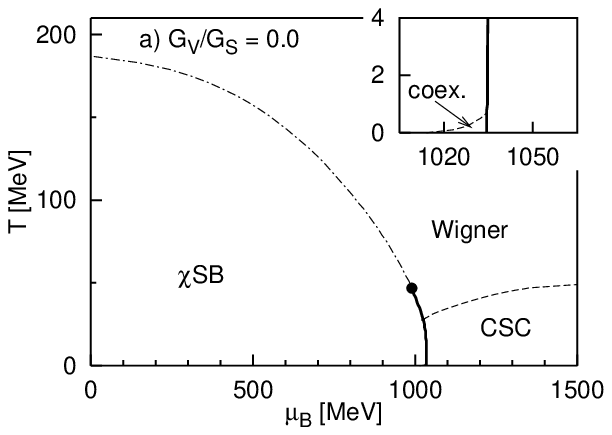}
\includegraphics[scale=1]{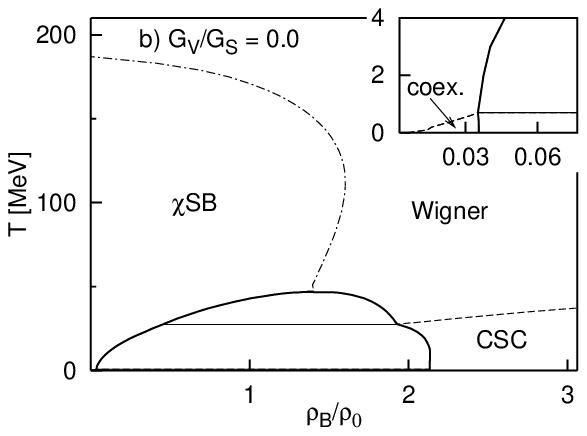}
\caption{
The phase diagrams in $T$-$\mu_B$ (a) and $T$-$\rho_B$ (b) plane
with $G_{V}=0$.
The solid line represents 
the critical line of a first-order phase transition,
the dashed line a second-order transition 
and the dot-dashed line a crossover.}
\label{fig:1}
\end{center}
\end{figure}
In Fig.~\ref{fig:1}(a), 
the phase diagram in the $T$-$\mu_B$ plane is shown.
One can see  that 
there are  four different phases, i.e. the $\chi$SB phase,
the Wigner phase, the CSC phase,
and the coexisting phase of $\chi$SB and CS\footnote{
The chiral condensate has a finite value even in the CSC phase,
because of the finite current quark mass.
However, the chiral condensate vanishes 
in the CSC phase in the chiral limit.
On the other hand, 
there exists a region  in which 
$\Delta$ becomes finite in the $\chi$SB phase even in the chiral limit.
We call this phase the coexisting phase.
}; 
as seen from the upper small 
panel, which is an enlargement of the part around
the solid line near $T=0$.
The figure also shows that the chiral transition 
is first-order at low temperatures 
and there is an endpoint at finite $T$ and $\mu_B$.
The phase transition from the CSC to the Wigner phase is second 
order in our model.
These features are 
qualitatively the same as that in \cite{ref:BR} 
except for the existence of the coexisting phase.
We have checked that 
the appearance of the coexisting phase with vanishing $G_V$ 
is sensitive to the parameters;
if a slightly larger $G_S$ is used, the coexisting phase disappears.

The corresponding phase diagram in the $T$-$\rho_B$ plane 
is shown in Fig.~\ref{fig:1}(b).
\begin{figure}[t]
\begin{center}
\includegraphics[scale=1]{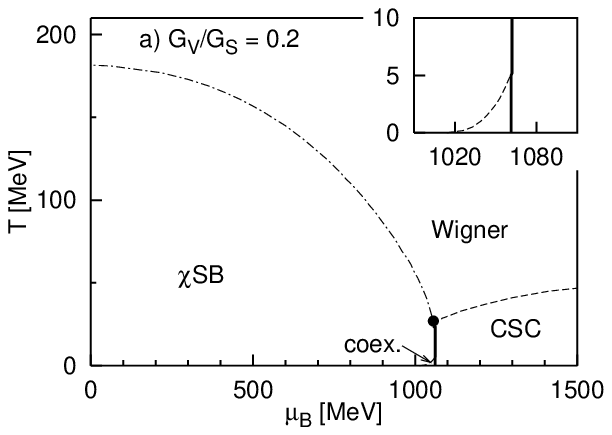}
\includegraphics[scale=1]{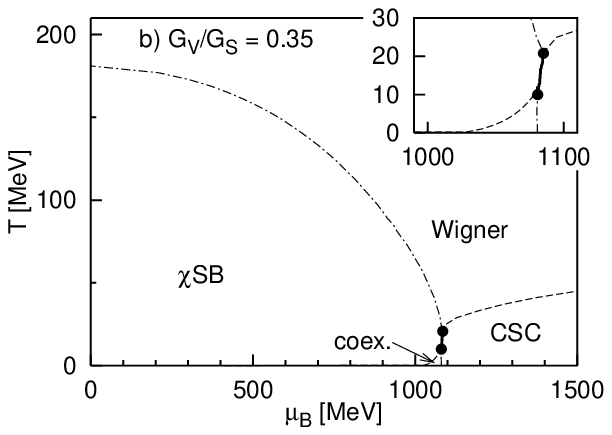}\\
\includegraphics[scale=1]{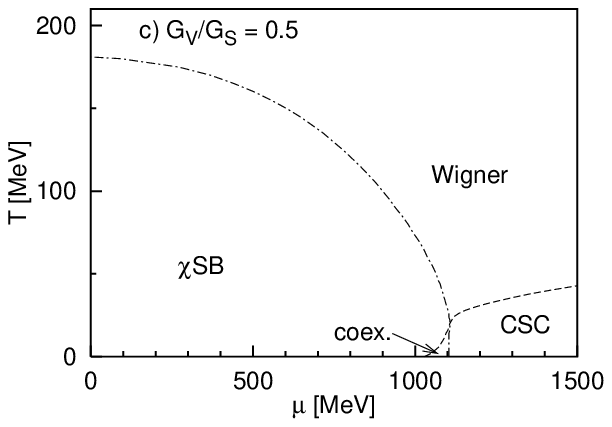}
\includegraphics[scale=1]{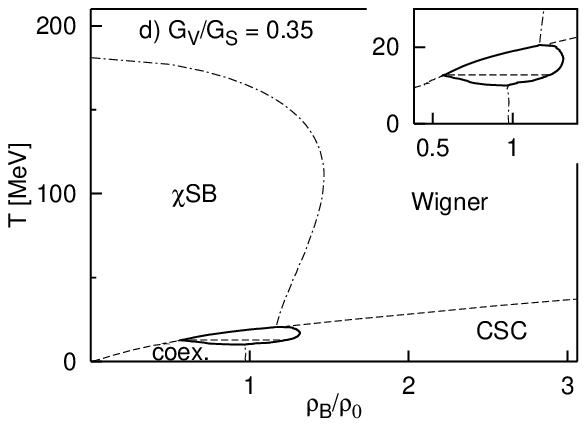}
\caption{
The phase diagrams with $G_V=0.2$ (a),$0.35$ (b), 
and $0.5$ (c) in the $T$-$\mu_B$ plane.
The phase diagram for $G_V=0.35$ in the $T$-$\rho_B$ plane 
is also shown in (d).}
\label{fig:2}
\end{center}
\end{figure}
\\

Now, we switch on the vector interaction
and discuss the effects of $G_V$.
In Fig.~\ref{fig:2}(a)-(c),
we show the phase diagram with $G_V= 0.2$, $0.35$, and $0.5$ 
in the $T$-$\mu_B$ plane.
From these figures, 
one can see that the phase structure is strongly affeted by $G_V$
especially near the critical line between the $\chi$SB and CSC phases:

\noindent
(1)
The endpoint of the first-order transition moves
toward a lower temperature as $G_V$ is increased,
and disappears eventually (Fig.~\ref{fig:2}(c)).
One can also see that
the vector coupling postpones the chiral restoration
toward larger $\mu_B$ at low temperatures.

\noindent
(2)
The region of the coexisting phase becomes broader 
with finite $G_V$.
We have checked that even with the large
$G_s$ as used in \cite{ref:BR}, 
the coexisting phase comes to exist as 
 $G_V$ is sufficiently increased.
Intuitively, this is because
the chiral restoration is shifted toward larger $\mu_B$ 
as $G_V$ is increased, then
the system can have a large Fermi surface even with 
the large consitutent quark mass $M$ owing to the large chiral
condensate;
the larger the Fermi surface, the larger the diquark condensate.

\noindent
(3)
In Fig.~\ref{fig:2}(b),
there appear two endpoints at both sides of the critical line 
of the first order transition.
In our calculation, such two-endpoint structure
gets to exist in a range of  $0.33\le G_V/G_S \le 0.38$.

It is also found that the thermodynamic potential
has a shallow minimum in a wide region of the $M_D$-$\Delta $ plane
near the two endpoints,
which  implies the large fluctuations of the chiral and 
diquark condensates.
The underlying mechanism to make the first order transition
to the second order in the low $T$ region is understood as
follows:
First of all, an enhanced coexisting phase is realized
due to $G_V$. Then in the high $\mu$ region,
the chiral condensate is suppressed owing to the CSC that 
makes the Fermi surface diffused. This works 
to weaken the phase transition as $T$ does
with the Fermi-Dirac distribution function.
%The numerical results suggest that 
%the endpoint at the lower temperature side emerges
%from the interplay between the $\chi$SB and CSC phases,
%i.e., the two-endpoint structure may not appear 
%if we do not consider the diquark interaction.

\section{Summary}

We have investigated  effects of the 
vector interaction (VI) on the chiral and color superconducting 
phase transitions at finite density and temperature.
We have shown that VI
enhances the interplay  between the $\chi$SB and CS phases
and that the phase structure is strongly 
affected by VI especially near the critical line between the 
$\chi$SB and  CS phases.

Although our analysis is based on a simple model,
the nontrivial interplay between the $\chi$SB and CSC phases
induced by VI 
is expected to be a universal phenomenon
and should
be confirmed and further studied 
with more realistic models including the random matrix model 
and on the lattice QCD.
As a future task, the color neutrality condition (CNC)
 should be taken into account;
although we believe that the present results 
for the effects of VI will not change
qualitatively with CNC incoporated,
the two-endpoint structure 
realized by a delicate interplay between $\chi$SB
and CS through VI
may disappear or persist in the mean-field approximation
we employed.

\end{document}